\documentclass[]{spie}  %>>> use for US letter paper
\usepackage{amsmath,amsfonts,amssymb}
\usepackage{graphicx}
\usepackage[colorlinks=true, allcolors=blue]{hyperref}

\newcommand{\aap}{Astronomy and Astrophysics}
\newcommand{\apj}{The Astrophysical Journal}
\newcommand{\mnras}{Monthly Notices of the Royal Astronomical Society}
\newcommand{\pasp}{Publications of the Astronomical Society of the Pacific}
\newcommand{\procspie}{Proceedings of the SPIE}
\newcommand{\apjl}{The Astrophysical Journal Letters}

\title{ESCAPE project: fundamental detection limits of JWST/NIRCam coronographic observations}

\author[1]{Godoy, N.}
\author[1]{Choquet, E.}
\author[1]{Altinier, L.}
\author[1]{Lau, A.}
\author[1,2]{Mayer, R.}
\author[1]{Vigan, A.}
\author[2]{Mary, D.}

\affil[1]{Aix Marseille Univ, CNRS, CNES, LAM, Marseille, France.}
\affil[2]{Université Côte d’Azur, Observatoire de la Côte d’Azur, CNRS, Laboratoire Lagrange.}

\authorinfo{N. Godoy.: E-mail: nicolas.godoy@lam.fr\\  E. Choquet: E-mail: elodie.choquet@lam.fr}

\begin{document} 
\maketitle

\begin{abstract}
 In this study, we explored the fundamental contrast limit of NIRCam coronagraphy observations, representing the achievable performance with post-processing techniques. This limit is influenced by photon noise and readout noise, with complex noise propagation through post-processing methods like principal component analysis. We employed two approaches: developing a formula based on simplified scenarios and using Markov Chain Monte Carlo (MCMC) methods, assuming Gaussian noise properties and uncorrelated pixel noise. Tested on datasets HIP\,65426, AF\,Lep, and HD\,114174, the MCMC method provided accurate but computationally intensive estimates. The analytical approach offered quick, reliable estimates closely matching MCMC results in simpler scenarios. Our findings showed the fundamental contrast curve is significantly deeper than the current achievable contrast limit obtained with post-processing techniques at shorter separations, being 10 times deeper at $0.5''$ and 4 times deeper at $1''$. At greater separations, increased exposure time improves sensitivity, with the transition between photon and readout noise dominance occurring between $2''$ and $3''$. We conclude the analytical approach is a reliable estimate of the fundamental contrast limit, offering a faster alternative to MCMC. These results emphasize the potential for greater sensitivity at shorter separations, highlighting the need for improved or new post-processing techniques to enhance JWST NIRCam sensitivity or contrast curve.
\end{abstract}

% Include a list of keywords after the abstract 
\keywords{JWST: NIRCam - High-contrast: infrared, coronagraphy - Post-processing: contrast limit, noise propagation - Direct imaging: Planet detection limits}

\section{INTRODUCTION} \label{sec:intro}  

Coronagraphic imaging has proved to be a powerful observational method that offers direct views of the outer regions of exoplanetary systems, which helps us to understand and characterize the atmosphere, demography, and formation pathways of exoplanets. Positioned at the L2 Lagrange point, the newly operational James Webb Space Telescope (\texttt{JWST}) promises significant advancements in the direct imaging of exoplanetary systems. This space-based observatory maintains a remarkably stable orbit, surpassing the capabilities of both the Hubble Space Telescope ([\citenum{Lallo+2006}]) and the most advanced ground-based adaptive optics systems ([\citenum{Beuzit+2019}]). The \texttt{JWST}’s post-launch image quality and stability exceed initial specifications ([\citenum{Rigby+2023}]), making it exceptionally suited for deep coronagraphic imaging tasks. Pre-launch models indicated its potential to detect exoplanets as small as $0.1\,\mathrm{M_{Jup}}$, an order of magnitude more sensitive than ground-based instruments, albeit at greater distances ([\citenum{Beichman+2010}]; [\citenum{Carter+2021}]). These theoretical predictions have been validated through on-sky results from the commissioning phase and Early Release Science (ERS) programs ([\citenum{Girard+2022}]; [\citenum{Boccaletti+2022}]; [\citenum{Carter+2023}]).

Our capability to eliminate contaminating starlight is crucial in determining the performance of coronagraphic imaging. Initially, the coronagraphic mask significantly reduces the starlight to contrast levels as low as $10^{-2}-10^{-3}$. Subsequently, further reduction of the remaining starlight that passes through the mask is achieved through advanced post-processing techniques. Currently, these post-processing methods primarily set the detection limits for objects at a short separation from the star.

A recommended observing strategy involves a precise sequence of actions to achieve optimal post-processing results in coronagraphic programs. Initially, observations of the science target, which could be with or without different roll angles, are immediately followed by those of a color-matched reference star. This procedure, referred to as Reference star Differential Imaging (RDI), is employed to facilitate the subtraction of starlight from the science images. The reference star is systematically dithered at the sub-pixel level around the mask center to account for \texttt{JWST} pointing errors, generating a comprehensive library of reference images. These reference stars images are subsequently processed using sophisticated algorithms, for example, Principal Component Analysis (PCA), to construct a model image that effectively minimizes starlight in the science data ([\citenum{Lajoie+2016}]; [\citenum{Soummer+2012}]).

However, the RDI+PCA approach is not without limitations. It leaves residual starlight at a contrast level of approximately $10^{-4}$ to $10^{-5}$ near the star. These residuals, characterized by azimuthal variations and known as speckle noise, impede the detection of exoplanets within $1''-2''$ from the star. This speckle noise, more accurately described as contamination, is a significant factor that cannot be mitigated by extending observation times, unlike other noise sources such as readout noise that predominate at greater separations. Consequently, the effectiveness of the post-processing phase in eliminating starlight residuals is crucial, as it sets a fixed contrast limit.

In coronagraphic images, we can distinguish between two detection regimes. The first regime, which occurs at short separations from the star, is dominated by speckles. This regime is defined by our capacity to subtract starlight and its radial statistics are determined by the combined performance of the telescope, coronagraph, and post-processing algorithm. Improving detection limits in this regime relies on the development of more effective starlight subtraction techniques. The second regime, present at larger separations, is dominated by regular noise. This noise includes stellar and thermal Poisson noise, as well as detector readout noise, and is constrained by classical imaging noise limits. Enhancing detection in this regime can be achieved by optimizing the exposure architecture and increasing the total exposure time.

In the context of the \texttt{ESCAPE} project ([\citenum{ChoquetSPIE2024}]), we analyze two detection methods by estimating the fundamental contrast limits of several JWST-NIRCam coronagraphy programs, breaking down the different noise sources. These limits help evaluate the performance of the RDI+PCA subtraction method and its potential improvements at short separations. We also investigate the separation distance at which the transition between speckle noise and readout noise occurs, and whether this transition depends on exposure time. We propose two methods for estimating the fundamental noise limit. In Section\,\ref{sec:definitions}, we introduce key concepts related to the fundamental contrast limit. In Section\,\ref{sec:approachs}, we present the ``Analytical approach'' and ``MCMC approach''. We apply these methods to three NIRCam datasets in Section\,\ref{sec:results}, and in Section\,\ref{sec:results+disc}, we present and discuss the main findings. Finally, in Section\,\ref{sec:concl}, we summarize our results and analysis.

\section{GENERAL FRAMEWORK AND NOTATIONS}\label{sec:definitions}

\subsection{Fundamental noise sources and notation}

\sloppy
In direct imaging, the scientific observing sequence is composed of a series of science integrations, for which we situate the science target behind the coronagraph to attenuate the starlight. Each science target integration, $S_{i}$, is affected by three noise sources: photon noise, readout noise, and flat noise. We note as $\sigma_{ph,S_{i}}$, $\sigma_{ron,S_{i}}$, and $\sigma_{flat,S_{i}}$ the standard deviation of each of these noises. They are evaluated by the \texttt{JWST} data processing pipeline\footnote{\url{https://jwst-docs.stsci.edu/jwst-mid-infrared-instrument/miri-observing-strategies/miri-coronagraphic-recommended-strategies}} and can be retrieved from the calibrated fits file extensions in MJy/sr, along with the image $S_{i}$ in the same units. Also provided in these files are the standard deviation of the total noise, $\sigma_{tot,S_{i}}$, which is the quadratic sum of the standard deviation of these three noise sources,
\fussy

\begin{equation}
    \sigma_{tot,S_{i}} = \sqrt{ \sigma_{ph,S_{i}}^2 + \sigma_{ron,S_{i}}^2 \sigma_{flat,S_{i}}^2 }
\end{equation}\label{eq:noises}

In the case of NIRCam coronagraphic observations, the science observing sequence dataset typically contains N integrations, each obtained with the telescope at an aperture angle $\theta_{i}$ from North. The derotated-combined raw image S can be obtained by:

\begin{equation}
    S = \frac{1}{N} \sum_{i=1}^{N} Rot_{\theta_{i}}(S_{i})
\end{equation}\label{eq:rot}

and the standard deviation of the corresponding noise sources of $S$ is:

\begin{equation}
    \sigma_{x,S} = \frac{1}{N} \sqrt{ \sum_{i=1}^{N} Rot_{\theta_{i}}(\sigma_{x,S_{i}}^2) } 
\end{equation}\label{eq:rot_noise}

with $x$ standing for the different noise sources ($ph$, $ron$, $flat$, and $tot$).

The starlight in these raw integrations is subtracted to improve the detection limits using an empirical PSF model. These models are computed from M reference star integrations, $R_{j}$ (reference differential imaging technique). As these are also affected by the same noise sources (noted as $\sigma_{ph,R_{j}}$, $\sigma_{ron,R_{j}}$, $\sigma_{flat,R_{j}}$, and $\sigma_{tot,R_{j}}$). The effect of the starlight subtraction is reflected in a decrease in the fundamental noise contribution in the starlight-subtracted images. 

Hence, we note $RES_{i}$ the starlight subtracted science integration, and $f$ the function that computes the PSF model:

\begin{equation}
    RES_{i} = S_{i} - f(R_{1}, ..., R_{M})
\end{equation}\label{eq:star-sub}

The standard deviation of the corresponding noise source is then:

\begin{equation}
    \sigma_{x,RES_{i}} = \sqrt{ \sigma_{x,S_{i}}^2 + g^2(R_{1},...,R_{N},\sigma_{x,R_{1},...,\sigma_{x,R_{M}}}) } 
\end{equation}\label{eq:res_star-ref}

with $g$ a function yielding positive values and corresponds to the noise propagation through the function $f$.

For instance, the classical subtraction of a single reference star image, of the same luminosity and exposure time as the science target would give:

\begin{equation}
    RES_{i} = S_{i} - R_{i}
\end{equation}\label{eq:simple_case}

and the standard deviation:

\begin{equation}
    \sigma_{x,RES_{i}} = \sqrt{\sigma_{x,S_{i}}^2 + \sigma_{x,R_{i}}^2}
\end{equation}\label{eq:noise_prop}

More generally, classical differential imaging uses as function $f$ a linear combination of the reference integrations, scaled to the science integration flux. The function $g$ that estimates the added noise on the residual image is thus easily computed as:

\begin{equation}
    g = \sqrt{ \sum_{k=1}^{M} [f'(R_{1},...,R_{M} | R_{k})\sigma_{R_{k}}]^2  }
\end{equation}\label{eq:cADI-noise}

\subsection{The case of PCA starlight subtraction}

On the other hand, starlight subtraction using Principal Component Analysis involves the calculation of the inverse covariance matrix of the reference star integrations, which is a non-linear operation for the function $f$. In addition, the science image (and its noise contributions) is projected onto the PCA modes of the reference images (and their filtered noise contributions). Both effects yield a non-trivial function $g$; in other words, it is unclear how the fundamental noise sources propagate through the PCA subtraction. In practice, quantitatively it is expected that PCA filters some of the reference frame noise contributions, in particular when selecting a low number of modes for the projection.

In order to estimate the fundamental noise limits in starlight-subtracted data despite this difficulty, we explore two approaches. In the first approach (``Analytical propagation''), we use several different assumptions on the effect of PCA on noise propagation to use linear models $f$, as in classical subtraction methods. In the second approach (``MCMC''), we estimate the noise on the PSF-subtracted images using Monte Carlo simulations. We describe both approaches in Section\,\ref{sec:approachs}.

\subsection{Contrast limits calculation}\label{contrast_curve}

The contrast limit is the limit for which our observations are sensitive to detect an astrophysical source, and it is calculated in the final, stacked residual image (e.g., eq.\,\ref{eq:star-sub}). The contrast limit can be estimated using different approaches (see [\citenum{Cantalloube+2021}] for a more complete and detailed summary). For this work, we use the ``classical'' approach described as follows. We take an annulus at $r\pm\Delta r$, being $\Delta r$ half of the resolution element. Then we take the standard deviation and the mean values in each annulus. We use these values to estimate the $5\sigma_{r}$ flux using the student-t statistics counting for low-number statistics ([\citenum{Mawet+2014}]). The contrast limit is defined as:  

\begin{equation}\label{eq:contrast}
C_{r} = \frac{5\sigma_{r} - m_{r}}{Th \cdot F_{star}} ,
\end{equation}

with $C_{r}$ the contrast at a separation of $r$, $5\sigma_{r}$ the \texttt{FPF} from the student-t distribution, $m_{r}$ the mean value in the annulus, $Th$ the throughput correction related to the over-subtraction in the post-processing method, and $F_{star}$ the star flux in the observed band/wavelength. 

In the case of the noise propagation, the ``noise'' contrast limit (or intrinsic contrast limit), can be calculated similarly, considering the noise nature of the stacked residual image. Here we assume we already have the noise image related to the stacked of residual frames. Since the image is already the noise, the mean value corresponds to the noise level (i.e., the equivalent to the standard deviation in the science frame in eq.\,\ref{eq:contrast}). Thus, the eq.\,\ref{eq:contrast} can be written as:

\begin{equation}\label{eq:contrast_noise}
C_{N,r} = \frac{5M_{r} - 0}{Th \cdot F_{star}} ,
\end{equation}

with $5M_{r}$ the \texttt{FPF} from the student-t distribution using the mean instead of the standard deviation as input. The contrast limit is then calculated at all angular separation with respect to the observed star.

\section{ESTIMATING UNCERTAINTIES IN POST-PROCESSED DATA}\label{sec:approachs}

We usually used different post-processing techniques to suppress the starlight, for which the intrinsic noise limit could be challenging to estimate. For example, principal component analysis works with covariance matrix and coefficients which could have a non-linear effect on noise propagation. Our goal is to find a method or approach that gives us the best approach to this fundamental contrast limit. The noise propagation can be bordered using two different points of view. The first one consists of finding a simple formula that allows us to estimate the limit faster and easily. The second consists of propagating the noise through the post-processing method using Monte Carlo, which requires more computing capacity and time but could be more accurate. We are aboard both below.

\subsection{Analytical Approach}\label{subsec:formulas}

We are working in this case with reference differential imaging and angular differential imaging with two different angle orientations. To simplify this, we take the basic and simplest case when we linearly combine the reference frames to reproduce the science, stellar coronagraphic PSF (classical differential imaging). We divided this approach into different cases, starting with the general formula and then simplifying it assuming different scenarios: 

\textbf{General case:} we assume that the reference ($R$) star is as bright as $D$ times the science ($S$) star. We estimate $D$ using the exposure times ($T$), so $D=T_{S}/ T_{R}$. This allows us to match the brightness of the reference star with the science star\footnote{Note that if we use the dither number instead of the exposure time, we can equalize the noise level at larger separations from the star (i.e., read-out-noise regimen)}. Then, we have $R=D\times S$. For the observation, we have $N$ science integrations (including the orientation angle) and $M$ reference frames (including the dithering). We can use $f$ in eq.\,\ref{eq:star-sub} as a linear reference frame combination to model each science integration $S_{i}$. Following eq.\,\ref{eq:star-sub}, the $RES$ can be written as follows:

\begin{equation}\label{eq:res_simple}
RES = \sum_{i=1}^{N}\frac{\bar{S_{i}}}{N} \, ,
\end{equation}

with $\bar{S_{i}}$ being the $i^{th}$ science integration already subtracted by the corresponding frame model using all the reference frames, and is defined as:

\begin{equation}\label{eq:Sbar}
\bar{S_{i}} = S_{i} - \sum_{j=1}^{M}\frac{\alpha_{j,i}R_{j}}{D} \, ,
\end{equation}

$\alpha_{j,i}$ refers to the coefficients we multiply to each reference frame to create our PSF model. The second term in the equation corresponds to $f$ in eq.\,\ref{eq:star-sub}. The normalization factor $D$ is used to equalize the flux in each reference frame to the science ones. Then, the sum of all $\alpha_{j}$ per PSF model is 1. When we also consider the different rolls angles used ($L$ rolls, $\theta_{1},...,\theta_{L}$), the total number of science frames is then defined as $N=\sum_{k=1}^{L}N_{k}$, being $N_{k}$ the number of frames on each roll angle. The general formula for the residuals is defined as:

\begin{equation}\label{eq:res_general}
RES = \sum_{k=1}^{L}\, Rot_{\theta_{k}}\left( \sum_{i=1}^{N_{k}}\frac{1}{N} \left[  S_{k,i} - \sum_{j=1}^{M}\frac{\alpha_{k,j,i}R_{k,j}}{D} \right] \right)\, ,
\end{equation}

with $S_{k,i}$ being each science integration at each respective roll angle, $Rot_{\theta_{k}}()$ the rotation function applied to a specific roll angle, and $R_{k,j}$ refers to each reference frame used for each PSF model of $S_{k,i}$. %for which we rotated to match the respective science frame.  

Now we consider the different noise contributions: photon, readout, and flat noises. The combination of all of them per frame is defined in eq.\,\ref{eq:noises} and referred to here as $\sigma_{ERR_{S}}$. Applying formulas \ref{eq:noise_prop} in \ref{eq:Sbar} we can obtain:

\begin{equation}\label{eq:err_Sbar}
\bar{\sigma}_{\bar{S_i}}^2 = \sigma_{S_{i}}^{2} + \sum_{j=1}^{M}\frac{\alpha_{j,i}^2\sigma_{R_{j}}^2}{D^2} \, ,
\end{equation}

Now propagating the last formula in equation \ref{eq:res_general} we have:

\begin{equation}\label{eq:err_general}
\sigma_{RES}^2 = \sum_{k=1}^{L} Rot_{\theta_{k}} \left( \sum_{i=1}^{N_{L}}\frac{1}{N^2} \left[ \sigma_{S_{k,i}}^{2} + \sum_{j=1}^{M}\frac{\alpha_{k,j,i}^2 \sigma_{R_{k,j}}^2}{D^2} \right] \right) \, ,
\end{equation}

We can simplify the equation \ref{eq:err_general} assuming different scenarios:

\textbf{Case 0:} we assume that the coronagraph can cancel perfectly the starlight, so we do not need to use the reference star, i.e., $\alpha_{j,i}=0$. We can re-write the equation\,\ref{eq:err_general} as follow:

\begin{equation}\label{eq:err_c0}
\sigma_{C0}^2 = \sum_{k=1}^{L} Rot_{\theta_k} \left( \sum_{i=1}^{N_{k}} \frac{\sigma_{S_{k,i}}^{2}}{N^2} \right) \, ,
\end{equation}

In this scenario, we assume we do not have coronographic starlight contamination. Therefore, we are underestimating the noise contribution. Consequently, we can consider the $\sigma_{C0}$ as a lower limit for the intrinsic contrast limit. 

\textbf{Case 1:} We assume that the PSF model has the same noise properties as the science integrations. Then:

\begin{equation}\label{eq:err_c1_a}
\sum_{j=1}^{M}\frac{\alpha_{k,j,i}^2 \sigma_{R_{k,j}}^2}{D^2}\, \equiv \, \sigma_{S_{k,i}}^2 ,
\end{equation}

The eq.\,\ref{eq:err_general} is re-written as: 

\begin{equation}\label{eq:err_c1}
\sigma_{C1}^2 = \, 2\,\times \,\sum_{k=1}^{L} Rot_{\theta_k} \left( \sum_{i=1}^{N_{k}} \frac{\sigma_{S_{k,i}}^{2}}{N^2} \right) \, ,
\end{equation}

In practice, this is equivalent to assuming that the noise is propagated with the same parameters/functions as the ones used for the PSF model. Then, the resulting PSF model noise will contribute with a similar noise level to the science integration. This should be approximately $2\times$ the case 0.

\textbf{Case 2:} We use only one reference frame per science frame. This is equivalent to assuming we can cancel each of the coronographic stellar PSFs perfectly with only one reference frame. Then, $\alpha_{j=1}=1$, with $D=T_{S}/T_{R}$. The equation \ref{eq:err_general} is then re-written as:

\begin{equation}\label{eq:err_c2}
\sigma_{C2}^2 = \sum_{k=1}^{L} Rot_{\theta_k} \left( \sum_{i=1}^{N_{k}} \frac{1}{N^2} \left[ \sigma_{S_{k,i}}^{2} + \frac{\sigma_{R_{k,j}}^2}{D^2} \right] \right) \, ,
\end{equation}

The noise level will be much higher than in case 1, and completely overestimated at large separations. This comes from two contributions: 1/ the low exposure time of the reference star, which means higher levels in the readout noise regimen, and 2/ we are not mean-combining the reference frames, so we are not reducing the associated noise. 

On the other hand, we are matching the flux of both stars ($D$), so the photonoise is not dominated by the reference star, but for both science and reference. We can consider this situation as a more realistic upper limit but still overestimated at larger separations.

\textbf{Case 3:} We use the eq.\,\ref{eq:err_general} with all the associated terms, assuming $\alpha_{k,j,i}=1/M$, i.e., mean combination of reference frames. We also assume $D=T_{S}/T_{R}$, so we are re-scaling the reference star flux to match with the science star flux. The eq.\,\ref{eq:err_general} is then:

\begin{equation}\label{eq:err_c3}
\sigma_{C3}^2 = \sum_{k=1}^{L} Rot_{\theta_k} \left( \sum_{i=1}^{N_{k}} \frac{1}{N^2} \left[ \sigma_{S_{k,i}}^{2} + \sum_{j=1}^{M}\frac{\sigma_{R_{k,j}}^2}{M^2D^2} \right] \right) \, ,
\end{equation}

With this scenario, we can obtain a contrast that will be higher than in case 0 and lower than in cases 1 and 2. This is because we use the average of $M$ frames instead of one single reference frame (decrease the noise contribution). Also, we are using the reference star noise which is not necessarily equal to the science frame noise. Indeed, the science noise will be combined with a lower noise coming from the average of all reference frames. We are weighting the noise of the reference frames with the factor $D=T_{S}/T_{R}$ too. We consider this approach the closest to the real contrast limit at larger separations, for which we assume a linear combination of the noise terms. At shorter separations, we could be underestimating the contribution of the reference photonoise.

\textbf{Case 4:} we assume that the $N$ science ($S$) and $M$ reference ($R$) frames have equal exposure time, brightness, and magnitude ($R=1\times S$ or $D=1$). Also, we assume a linear and equal-weighted combination of the $M$ reference frames to create the model of each science frame, i.e. $\sum_{j}\alpha_{i,j} = 1$ with $\alpha_{i,j}=\alpha = 1/M$. This is, case 3 using $D=1$. The final residual image is then averaged using all the residual frames. The equation \ref{eq:err_c3} can be written as:

\begin{equation}\label{eq:err_c4}
\sigma_{C4}^2 = \sum_{k=1}^{L} Rot_{\theta_k} \left( \sum_{i=1}^{N_{k}} \frac{1}{N^2} \left[ \sigma_{S_{k,i}}^{2} + \sum_{j=1}^{M}\frac{\sigma_{R_{k,j,i}}^2}{M^2} \right] \right) \, ,
\end{equation}

This scenario is not realistic since we are assuming the same brightness between science and reference stars, the same exposure time, and a factor $\alpha_{j,i}$ equal for all the reference frames. This depends on the strategy adopted, but usually, the reference star is brighter than the science star. This will give us a contrast much higher than the real one, fully dominated by the reference star noise sources. Note that all the integration (science and reference), are already normalized by the exposure time. Then, the photons and background are higher for the reference which has a lower exposure time. In practice, this means a higher level in the readout noise regimen. This scenario is an extreme, unrealistic upper limit.

\subsection{MCMC Approach}\label{subsec:MCMC}

The MCMC approach consists of using directly the post-processing, in our case $\mathrm{KLIP}$, to propagate the noise. Each of the integrations for the science and reference stars has its associated noise term. We assume, for each pixel, a Gaussian distribution with the mean and standard deviation being the integration value and noise (i.e., the $\mathrm{RES}$, $\mathrm{RON}$, $\mathrm{Poisson}$, or $\mathrm{Flat}$ in the \texttt{JWST} products). We assume the non-correlation scenario of pixels near each other. Then, we can compute $10^{\gamma}$ new datasets\footnote{Usually, $\gamma$ is $\gamma = 4$ to generate a considerable sample for statistical analysis.} to use as input in the post-processing technique. We generate $10^{\gamma}$ new stacked of residual frames, for which we estimate our representative stacked residual frame and the related noise by calculating the mean and standard deviation. Finally, we use the formulas\,\ref{eq:contrast} and\,\ref{eq:contrast_noise} to estimate the respective sensitivity and fundamental contrast limit. Since we are adding more noise to the images (from the Gaussian approach and the non-correlated pixel assumption), we estimate the normalization factor to match the sensitivity limit (contrast curve) obtained with the MCMC to match the one obtained when using \texttt{KLIP}.

\section{CONTRAST LIMITS FROM BOTH APPROACHES}\label{sec:results}

We tested both approaches in three different datasets observed with different filters: HIP\,65426, AF\,Lep, and HD\,114174. These datasets were taken from the \texttt{ERS} programs \texttt{ERS1386}, \texttt{ERS4558}, and \texttt{ERS1441}. Table\,\ref{table:Obs_summary} summarizes the main observational setup for each target. We computed the contrast limit using \texttt{KLIP} with the maximum number of components to use as a reference in our analysis. The contrast limits for each target using both approaches are described below.

\begin{table*}[h]
\centering
\caption{Summary of the dataset/observations used in this study. }
  \begin{tabular}{ l c c c c c c c }
    \hline \hline
  Program & Target Name & Type & Filter &  $\mathrm{N_{Groups}}$ & $\mathrm{N_{Int}}$ &  $\mathrm{T_{Int}}$ & Total integrations   \\
   &   &   &   &  &  &  [sec] & \\
%       &            &                &             &    \\ 
    \hline
\texttt{ERS1386} & HIP\,65426 & SCI & \texttt{F444W} & 15 & 2  & 307 & 4  \\
        & HIP\,68245 & REF & \texttt{F444W} & 4  & 2  & 41  & 18 \\
\texttt{ERS4558} & AF\,Lep    & SCI & \texttt{F356W} & 5  & 35 & 26  & 70 \\
        & HD\,33093  & REF & \texttt{F356W} & 5  & 15 & 26  & 252 \\
\texttt{ERS1441}& HD\,114174 & SCI & \texttt{F335M} & 10 & 63 & 53  & 126 \\
        & HD\,111733 & REF & \texttt{F335M} & 10 & 7  & 53  & 63 \\
\hline
  \end{tabular}
\label{table:Obs_summary}
\end{table*}

\subsection{Formula approach estimates}\label{subsec:formula_est}

\begin{figure*}[h]
\centering
\includegraphics[width=7.75cm]{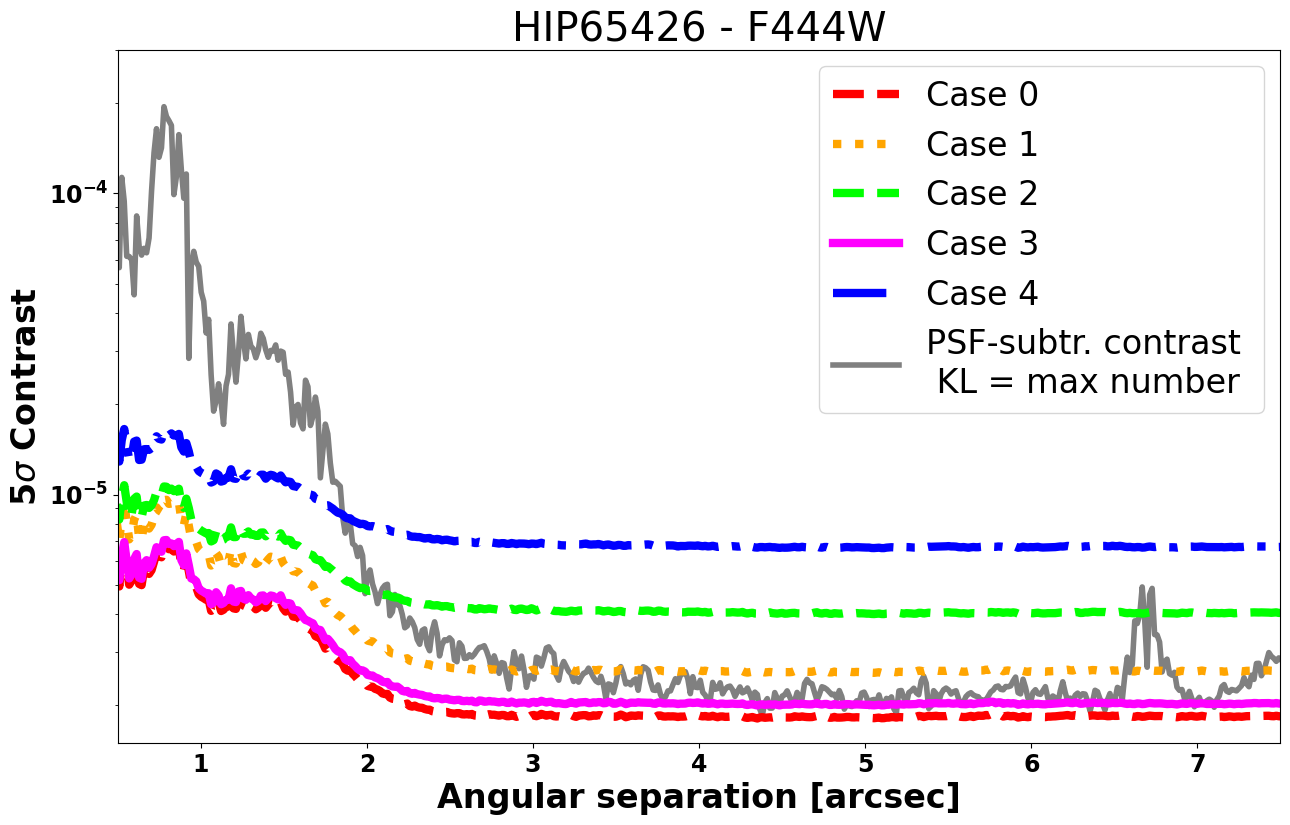}
\includegraphics[width=7.75cm]{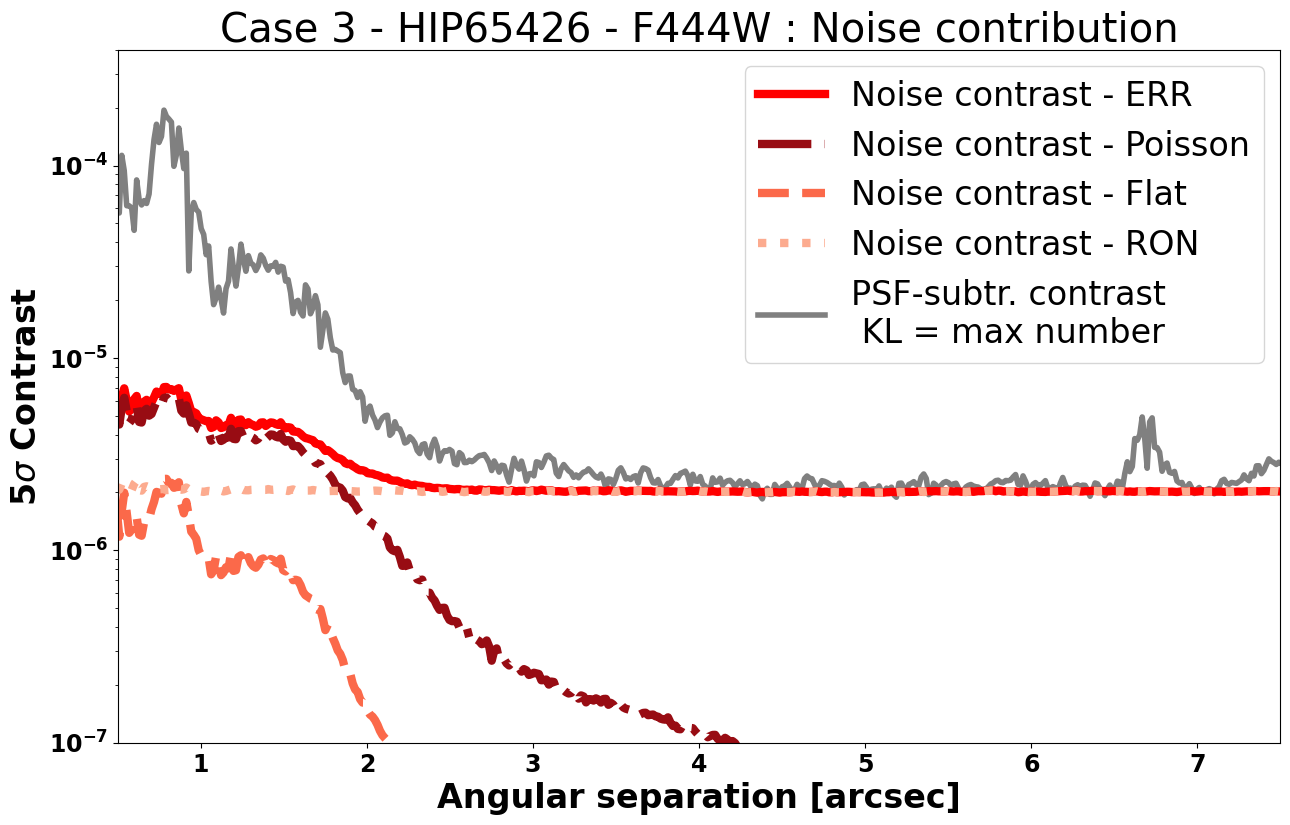}
\includegraphics[width=7.75cm]{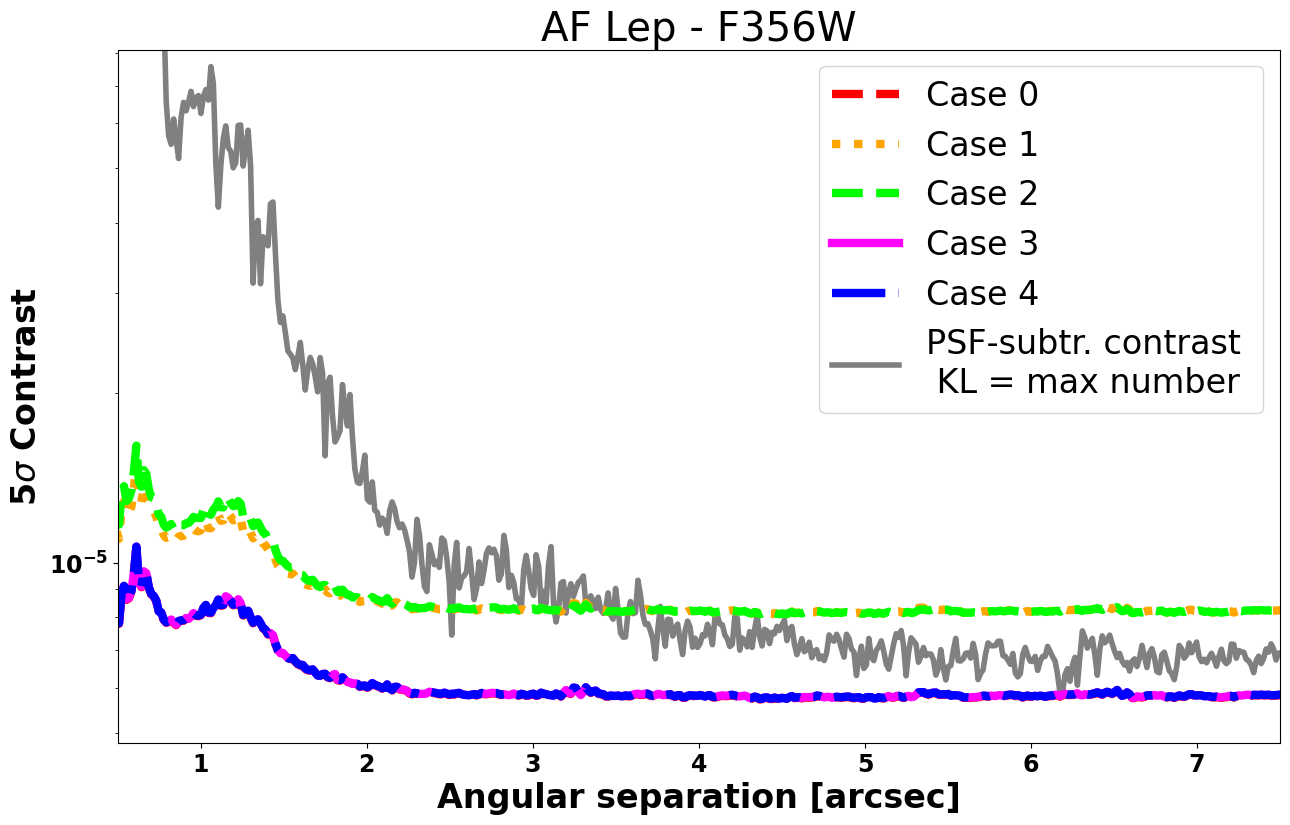}
\includegraphics[width=7.75cm]{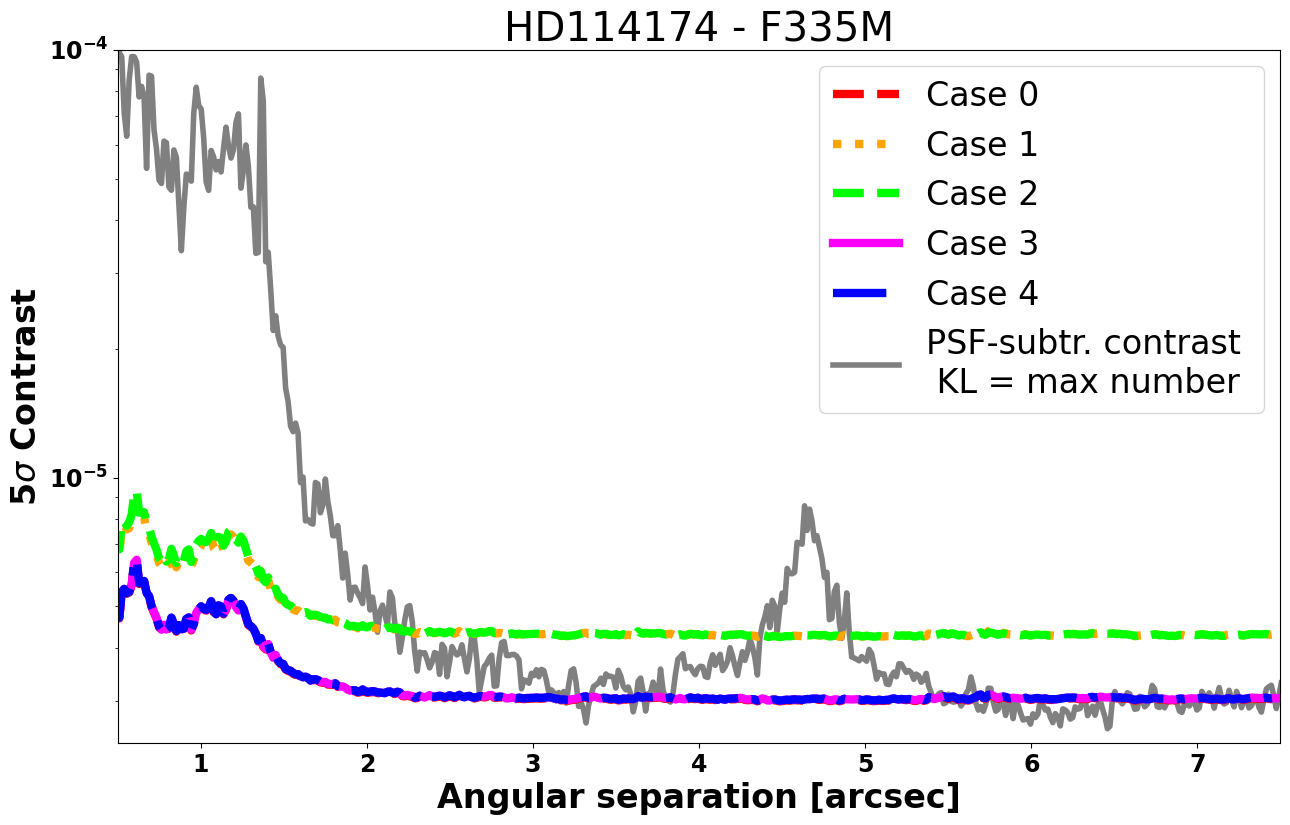}
\caption{ Contrast limits using the formula approach for HIP\,65426, AL\,Lep, and HD\,114174. \texttt{Top-Left}: the five different cases to obtain the fundamental contrast limits for HIP\,65426. Each of the colored line correspond to each case. \texttt{Top-Right}: the different contributions of the noises (Poisson noise, flat noise, and read-out noise) and the total noise for HIP\,65426 case 3. Each of the red lines corresponds to each noise. \texttt{Bottom-Left:} same as top-left but for AF\,Lep. \texttt{Bottom-Right:} same as top-left but for HD\,114174. The gray line in both figures corresponds to the contrast limit obtained using \texttt{KLIP}.}
\label{fig:all-3_formula}%
\end{figure*}

First, we computed the contrast limit for the three targets using the 5 different scenarios presented in Section\,\ref{subsec:formulas}. Figure\,\ref{fig:all-3_formula} shows the results using the formula approach presented in section\,\ref{subsec:formulas} for all five cases. In the top for HIP\,65426, and bottom for AF\,Lep and HD\,114174. For the last two targets, some of the cases match each other given the observing setup and configuration (e.g., the same number of groups, and exposure times, see Table\,\ref{table:Obs_summary}). For all three targets, the computational time was of a few minutes.

\subsection{MCMC approach estimates}\label{subsec:MCMC_est}

We simulate the sub-datasets for each \texttt{ERS} program: the science and reference observations. For each simulated dataset, we save the new science and reference frames, the PCA base, the PSF model, and the residual stacked image. Note that we repeated the simulation using each of the noise components alone, and the combination of all the noise sources ($4\times$ the sample size). For HIP\,65426 we simulated $7\times10^3$ new samples (reference and science frames). For AF\,Lep and H\,114174, $10^2$ and $1.5\times10^2$, respectively. The differences in the number of simulated samples come from the number of integrations for each program, being AF\,Le and HD\,114174 extremely large. For example, for HIP\,65426 we required less than 200\,Gb of memory and less than 6 hours to create and process 1\,000 simulated datasets\footnote{We tested the procedure simulating 300 datasets, then we simulated 700 new datasets. The last group was used in our analysis.} and to compute the final contrast curves. For HD\,114174 we required 5 days and more than 850\,Gb of memory to save the information (simulates samples, PCA, and residuals related to each noise source) of about 150 samples.

\begin{figure*}[h]
\centering
\includegraphics[width=7.75cm]{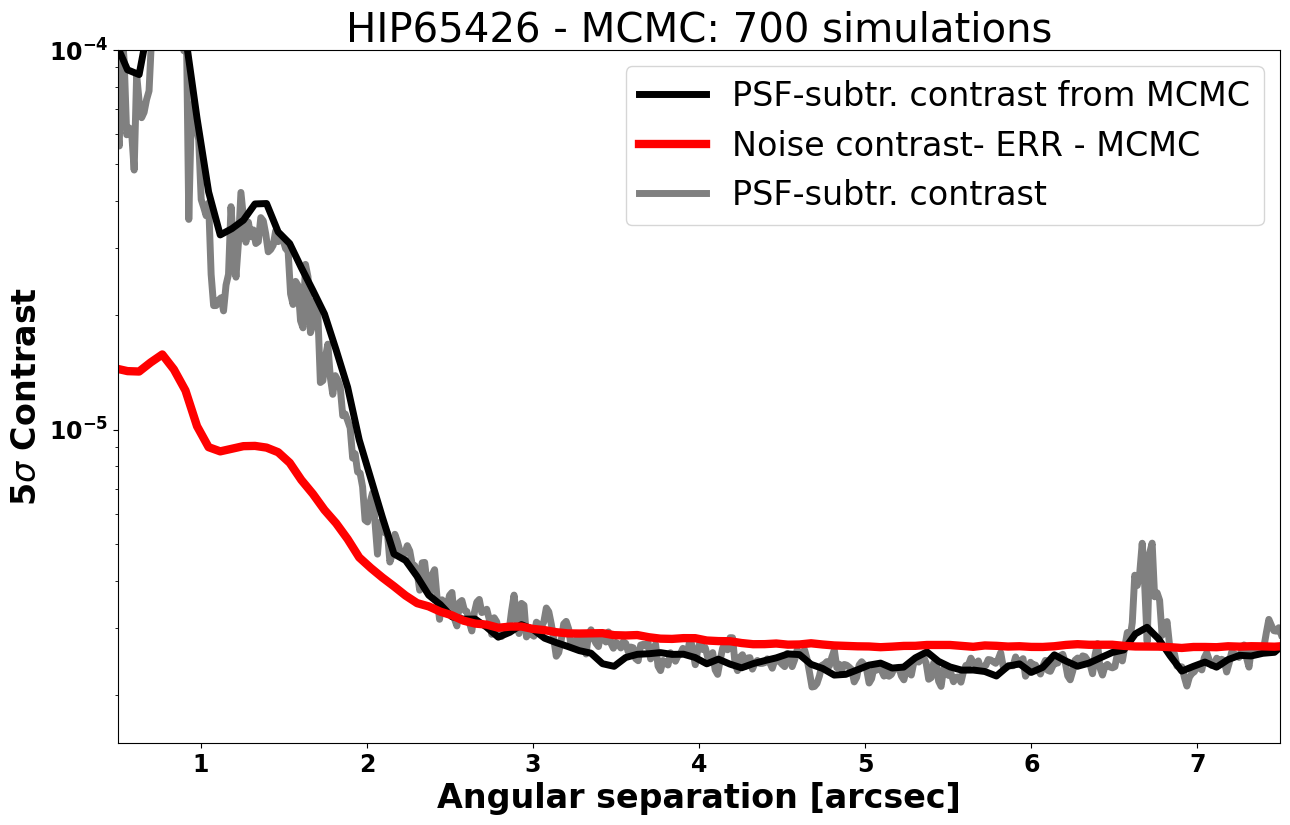}
\includegraphics[width=7.75cm]{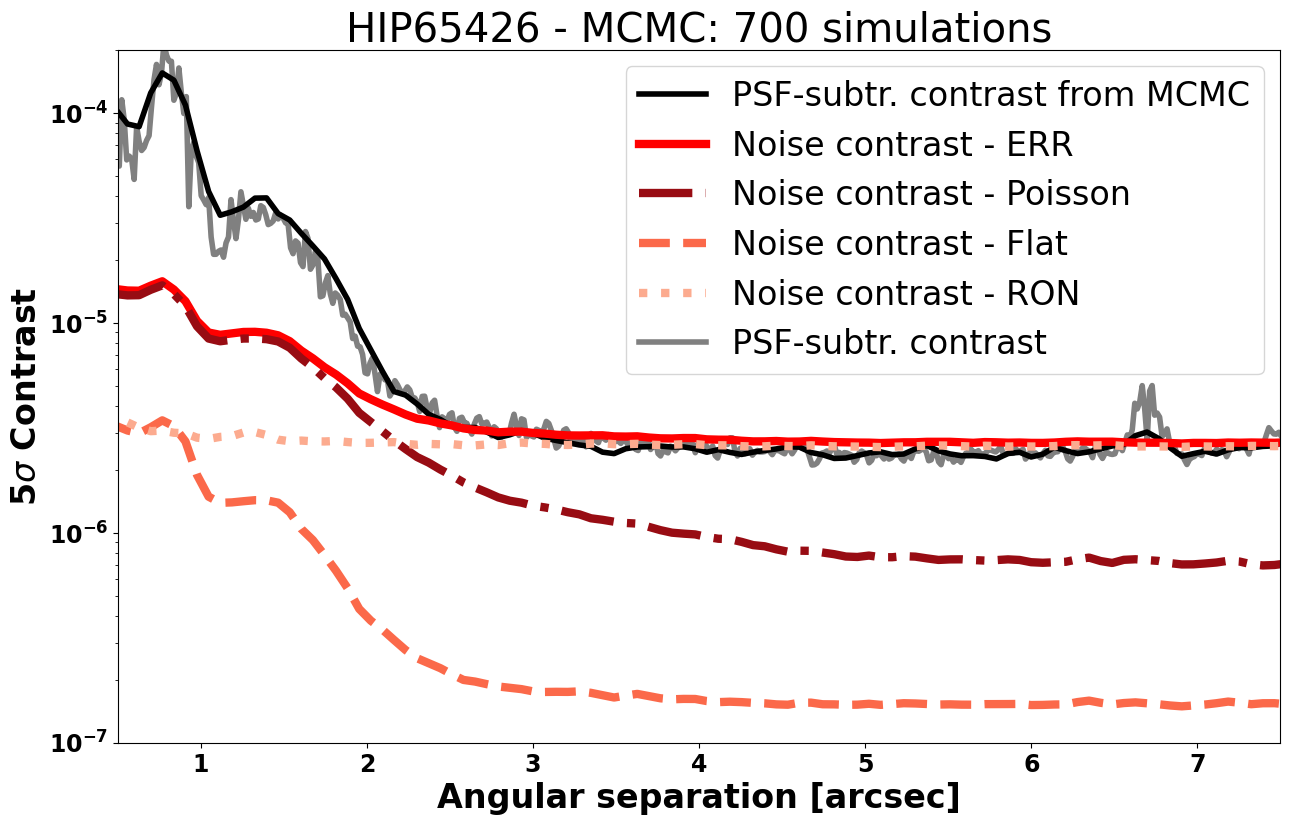}
\includegraphics[width=7.75cm]{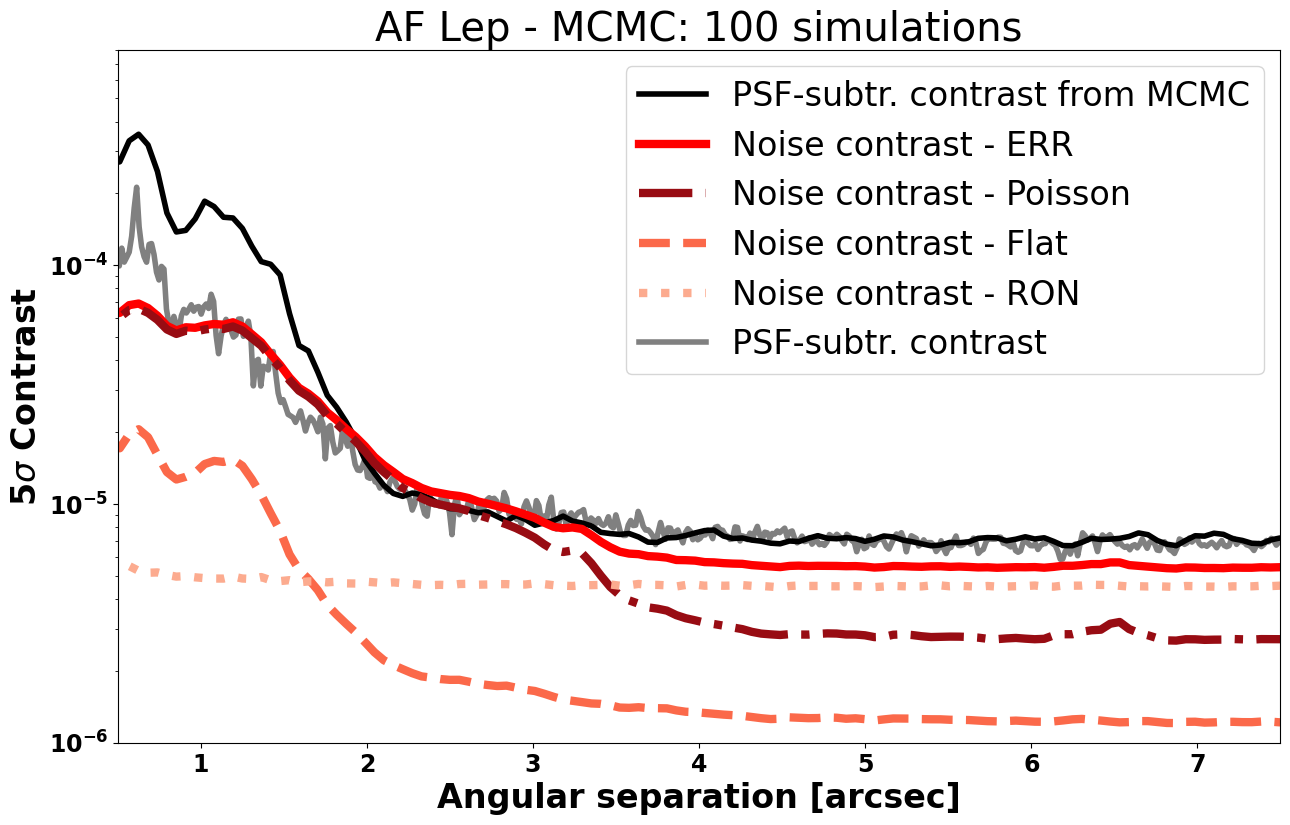}
\includegraphics[width=7.75cm]{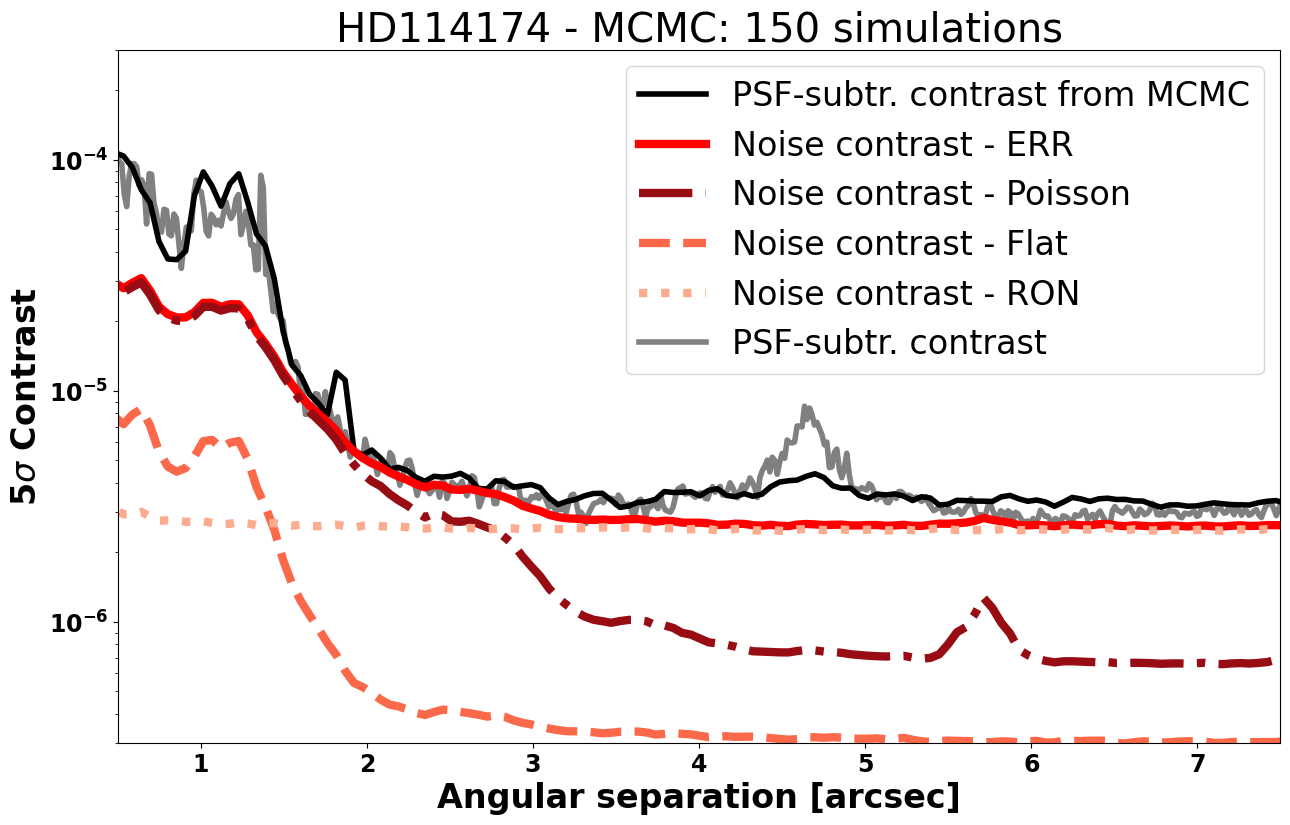}
\caption{ Contrast limits using the MCMC approach for HIP\,65426, AF\,Lep, and HD\,114174. We used 700 of the 1000 simulations in the case of HIP\,65426. \texttt{Top Left:} the contrast curve from \texttt{KLIP} (gray curve), the contrast curve from MCMC (black curve), and the noise curve or fundamental contrast limit (red curve) for HIP\,65426. \texttt{Top Right:} each of the noise components in the MCMC for HIP\,65426. \texttt{Bottom Left:} same as the figure in the top-right but for AF\,Lep. \texttt{Bottom Right:} same as the figure in the top-right but for HD\,114174.}
\label{fig:all-3_MCMC}%
\end{figure*}

Figure\,\ref{fig:all-3_MCMC} shows on the top-left side the MCMC contrast limit and fundamental sensitive (noise contrast limit) for HIP\,65426, and on the top-right the contribution of each noise source. The bottom side is the same as the top-right but for AF\,Lep (bottom-left), and HD\,114174 (bottom-right).

\section{RESULTS AND DISCUSSION}\label{sec:results+disc}

The two approaches (formula and MCMC) help us estimate how close we are to the fundamental contrast limit. The formula method is quick and efficient for determining this limit but requires testing and calibration. In contrast, the MCMC approach provides the most accurate noise propagation and fundamental contrast limit recovery in post-processed images, assuming certain noise properties. However, it demands significantly more computational power and memory. Additionally, this method does not account for pixel correlation, affecting sensitivity. Also, the statistical sample size is often too small for precise convergence due to computational constraints. However, for HIP\,65426, the sample size was sufficient to ensure MCMC convergence.

\begin{figure*}[h]
\centering
\includegraphics[width=7.75cm]{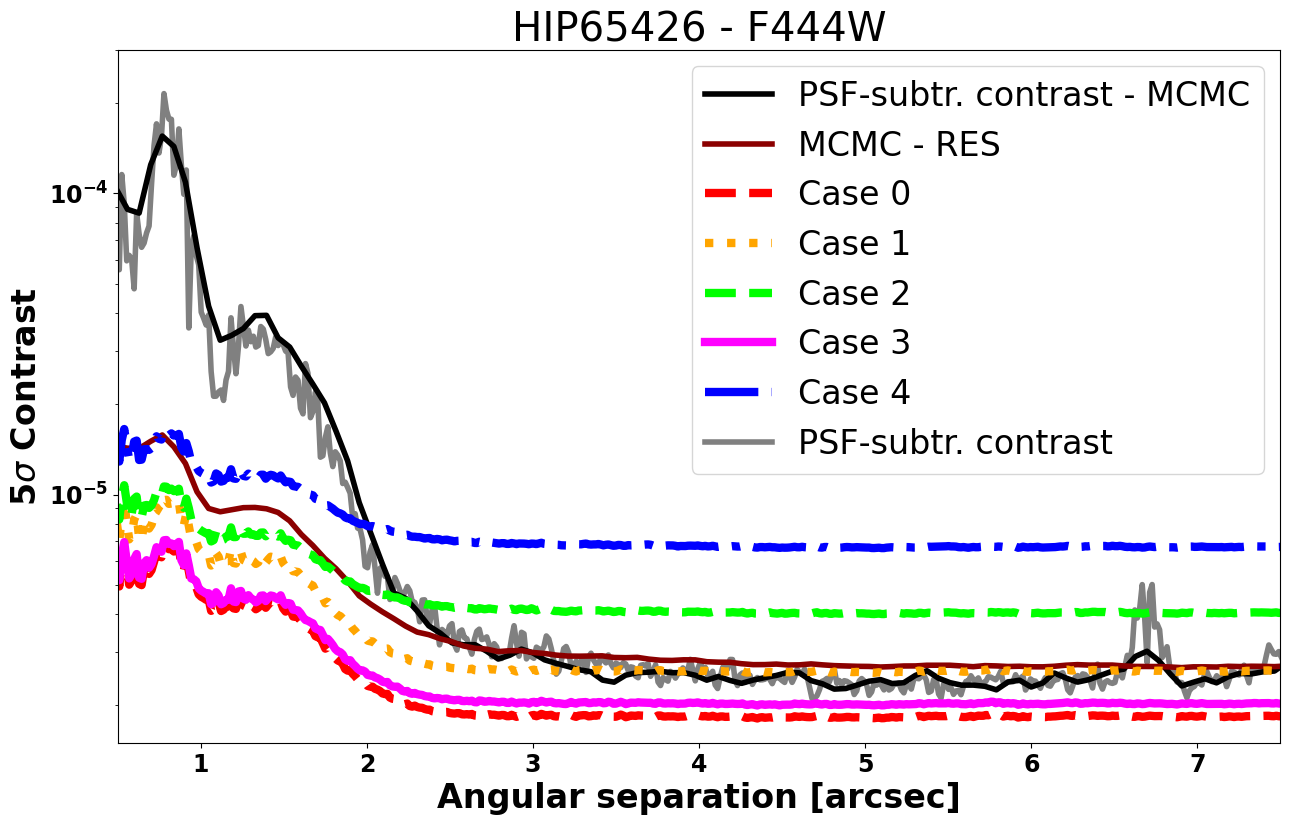}
\includegraphics[width=7.75cm]{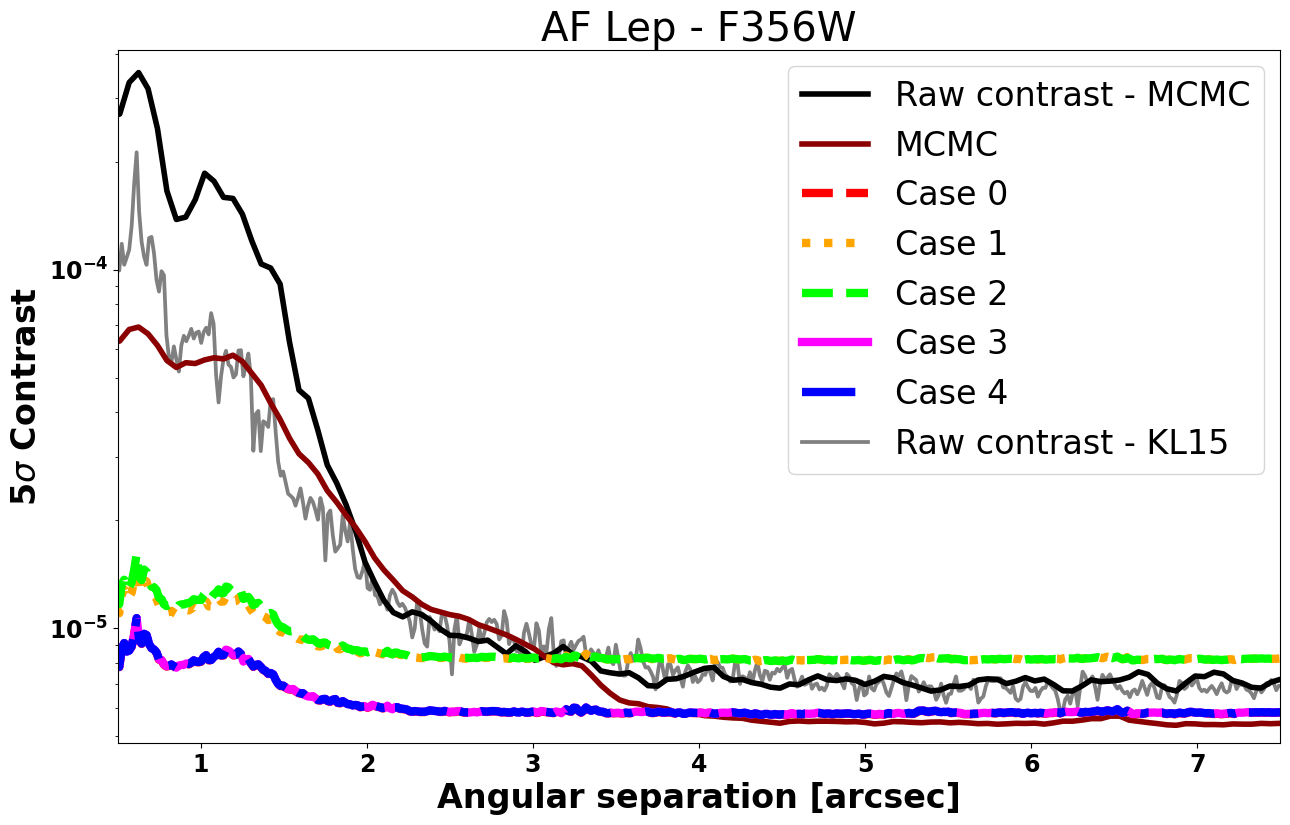}
\caption{Sensitivity and fundamental sensitivity for the formula and MCMC approach. \texttt{Left:} HIP\,65426. \texttt{Right:} AF\,Lep/ The different color lines correspond to the sensitivity for each different approach in the formula case and the one from MCMC. The gray curve corresponds to the contrast limit using \texttt{KLIP} and the black one using MCMC.}
\label{fig:comparison}%
\end{figure*}

Figure\,\ref{fig:comparison} compares the performance of various formulas and the MCMC method for HIP\,65426 (left) and AF Lep (right). We can highlight the following: \textbf{1/ Noise Matching}: The MCMC method aligns with the read-out noise for HIP\,65426 and performs even better for AF\,Lep due to the higher noise from shorter exposure times (in the case of AF\,Lep), enhancing contrast further from the central region. \textbf{2/ Sample Size Impact:} MCMC contrast closely matches the \texttt{KLIP} contrast with large sample sizes (1000. HIP\,65426), while for smaller samples (100, AF\,Lep), starlight is poorly subtracted, indicating more samples are necessary to accurately represent fundamental sensitivity. \textbf{3/ Sensitivity Improvement:} MCMC demonstrates significant sensitivity improvements in the photonoise regime, achieving a 10x improvement at $0.5''$ and 4x at $1''$. Enhanced post-processing techniques can further improve contrast limits. Increasing exposure time and group numbers can enhance readout noise regime contrast limits. HIP\,65426, HD\,114174, and AF\,Lep achieve contrast levels of $2.5 \times 10^{-5}$, $3 \times 10^{-5}$, and $7 \times 10^{-5}$, respectively, correlating with exposure time and group numbers (see Table\,\ref{table:Obs_summary}). \textbf{4/ Exposure Time and Noise Regions:} Increasing exposure time, and thereby the number of groups, reduces the photon noise-dominated region. For HIP\,65426, the transition from photon noise to readout noise occurs at $\sim 2.2''$, while for AF\,Lep, it is at $\sim 3''$, mainly due to differences in integration time and group numbers. \textbf{5/ Read-out Noise Matching between approaches:} For the readout noise-dominated region, case 1 closely matches MCMC, while case 3 provides a lower limit. In the photon noise regime, cases 1 and 2 differ from MCMC by a factor of 2. Considering MCMC might be slightly overestimated (due to not accounting for pixel correlation), case 1 is a good approximation, and case 3 serves as a lower-limit estimator. \textbf{6/ \texttt{KLIP} Noise Propagation:} \texttt{KLIP} propagates reference frames noise similarly to how it models the PSF of coronagraphic science frames, meaning it creates a model for both the stellar PSF and noise properties in the same way.

\section{Summary and conclusions}\label{sec:concl}

In this study, we explored the fundamental contrast limit of NIRCam coronagraphy observations, which represents the performance achievable with post-processing techniques. This limit is influenced by various noise sources, primarily photon noise and readout noise, but understanding how these noises propagate through post-processing is complex. For example, principal component analysis involves matrices and coefficients that do not linearly affect noise propagation. We studied the noise propagation in the framework of angular differential imaging with reference star differential imaging using KLIP for the post-processing technique.

To determine the fundamental contrast limit and understand noise propagation, we employed two approaches. The first approach involved developing a formula based on simplified scenarios, from considering only science observation noise to including all noise sources from both science and reference star observations. We used the simplest approach consisting of a linear combination of the reference star observations to reproduce the coronographic stellar PSF. The second approach used Markov Chain Monte Carlo (MCMC) methods, assuming Gaussian noise properties and uncorrelated pixel noise.

We tested both approaches on three datasets: HIP\,65426, AF\,Lep, and HD\,114174. The MCMC method provided accurate estimates but was computationally intensive and dependent on our assumptions. The analytical approach, though less precise, offered quick estimates and closely matched the MCMC results in simpler scenarios.

Our findings showed that the fundamental contrast curve is significantly deeper at shorter separations, being 10 times deeper at $0.5''$ and 4 times deeper at $1''$. This corresponds to the photon noise regimen, and it is limited to our capacity to subtract the starlight in post-processing techniques. At greater separations, we have the read-out noise regimen, for which increasing the exposure time we can improve the sensitivity. The transition between photon noise and readout noise dominance occurs between $2''$ and $3''$, largely depending on exposure time.

We conclude that the analytical approach is a reliable estimate of the fundamental contrast limit, offering a faster alternative to the MCMC method. We can obtain lower and upper limit estimations on the actual fundamental contrast curve. These results emphasize the potential for greater sensitivity at shorter separations, suggesting a need for improved or developed new post-processing techniques to enhance \texttt{JWST} NIRCam contrast curves.

\acknowledgments % equivalent to \section*{ACKNOWLEDGMENTS}       
 
This project is funded by the European Union (ERC, ESCAPE, project No 101044152). Views and opinions expressed are however those of the authors only and do not necessarily reflect those of the European Union or the European Research Council Executive Agency. Neither the European Union nor the granting authority can be held responsible for them.  

% References

 % bibliography data in report.bib
\bibliographystyle{spiebib} % makes bibtex use spiebib.bst

\end{document}